\documentclass[
	aps, 
	prd,  
	notitlepage, 
    floats, 
    floatfix, 
    onecolumn,
	amsmath, 
	amssymb, 
	amsfonts, 
	eqsecnum,
	superscriptaddress,
	showpacs, 
	showkeys,
	nofootinbib,
 	longbibliography,
]{revtex4-1}

\usepackage{graphicx}% Include figure files
\usepackage{dcolumn}% Align table columns on decimal point
\usepackage{bm}% bold math
\usepackage{color}

\def\comment#1{}

\begin{document}

\title{Detecting properties of echoes from inspiraling stage with ground-based detectors}% Force line breaks with \\
%\thanks{This work is supported by NSFC No. U1431120 and No.11273045  }%

\author{Xing-Yu Zhong}
\email{zxy@shao.ac.cn}
\affiliation{Shanghai Astronomical Observatory, Chinese Academy of Sciences, Shanghai 200030, China}
\affiliation{School of Astronomy and Space Science, University of Chinese Academy of Sciences, Beijing 100049, China}

\author{Wen-Biao Han}
\email{corresponding author: wbhan@shao.ac.cn}
\affiliation{School of Fundamental Physics and Mathematical Sciences, Hangzhou Institute for Advanced Study, UCAS, Hangzhou 310024, China}
\affiliation{Shanghai Astronomical Observatory, Shanghai, 200030, China}
\affiliation{School of Astronomy and Space Science, University of Chinese Academy of Sciences, Beijing 100049, China}
\affiliation{International Centre for Theoretical Physics Asia-Pacific, Beijing/Hangzhou, China}
\affiliation{Key Laboratory for Research in Galaxies and Cosmology, Shanghai Astronomical Observatory, Shanghai 200030, China}

\author{Ye Jiang}
\affiliation{Shanghai Astronomical Observatory, Chinese Academy of Sciences, Shanghai 200030, China}

\author{Ping Sheng}
\affiliation{Shanghai Astronomical Observatory, Chinese Academy of Sciences, Shanghai 200030, China}

\author{Shu-Cheng Yang}
\affiliation{Shanghai Astronomical Observatory, Chinese Academy of Sciences, Shanghai 200030, China}

\author{Chen Zhang}
\affiliation{Shanghai Astronomical Observatory, Chinese Academy of Sciences, Shanghai 200030, China}

\date{\today}% It is always \today, today,

\begin{abstract}
The nature of black holes is one of most exciting issues in gravitational physics.  If there is an exotic compact object as the compact as a black hole but without a horizon, gravitational wave echoes may be produced after the merger. 
In this work, we show that for extreme-mass-ratio binaries, even during the inspiraling phase of compact binary coalescence, the exists of hard surface of the exotic compact object will produce detectable signals on the gravitational waves. We predict that once the LIGO-Virgo-KAGRA, Einstein Telescope or Cosmic Explorer detect such kind of sources, our model shows that one can constrain the properties of surfaces of the compact objects in inspiraling stage better than the current level.
%\begin{description}
%\item[PACS numbers] 04.70.Bw, 04.80.Nn, 95.10.Fh
%\end{description}
\begin{description}
\item[PACS numbers]
04.70.Bw, 04.80.Nn, 95.10.Fh
%May be entered u\sing the \verb+\pacs{#1}+ command.

\end{description}
\end{abstract}
% \pacs{04.70.Bw, 04.80.Nn, 95.10.Fh}% PACS, the Physics and Astronomy
\maketitle
%\tableofcontents

%\tableofcontents
%{\it Accurate simulation of GW sources for mini-Taiji} by HAN Wen-Biao\\
%\email{wbhan@shao.ac.cn}

\section{introduction}
The updated gravitational wave (GW) catalog shows that more than 90 GW events have been detected by LIGO-Virgo-KAGRA (LVK). Most of these events are inferred as binary black holes or black hole - neutron star mergers \citep{gwcatalog}. These observations give us a unique opportunity to test the nature of black holes, especially the horizon which is a key task in the GW astronomy \citep{2017PhRvL.119i1101K, 2021arXiv211206861T}.  

Exotic Compact Objects (ECOs) are black hole like but horizonless compact objects other than a neutron star. The candidates of ECOs could be abundant, such as boson star, fluid star, gravastar, bubble, fuzzball, superspinars, and etc \citep{rubio18,cardoso19}. These ECOs are black holes like but without horizons, replaced by surfaces very close to horizons of the corresponding black holes. The surface locates at the Planck scale out side of the horizon, and new physics may hide. The radius of surface of ECO is $r_0 = (1+\epsilon) (M+\sqrt{M^2-a^2})$, where $a$ is the Kerr parameter. While  $\epsilon = 0$, we go back to black hole. For neutron stars, $\epsilon \sim \cal{O}$$(1)$.  For ECOs like as boson stars, we expect that $r_0 - r_{\rm h}$ is of the order of the Planck length $l_P$, in such case $\epsilon \sim 10^{-40} $ or even smaller. Assuming $\epsilon \ll 1$, one can test the BH paradigm in an agnostic way, or for testing the effects of quantum gravity \citep{cardoso19}. 

For black holes, the effective potential will reduce to zero after the light ring. However, for a horionless compact object, due to the hard surface, there is a potential well supports quasi-trapped, long-lived modes, which is called as gravitational wave echoes \citep{cardoso17}. Theoretically, people discussed the physical mechanism of echoes from ECOs \citep{echoSchw,duPRL18,chen2019instability}. Some of us calculated echoes from Kerr-like ECOs using the Teukolsky equation with reflecting boundary \citep{xin2019}. A few literature have tried to find echoes from the LIGO data , and give the parameter estimations on ECOs\citep{echoSearch1,echoSearch2,echoSearch3,echoSearch4,echoSearch5}.

All these attempts are searching the echoes directly after ringdown to constrain the exist of ECOs. From Fig. \ref{gwecho}), one can see that the echo signal is much weaker than the inspiral-merger-ringdown one. Therefore, it is difficult to constrain or detect echoes. However, due to the reflection property of the surface of ECO, the energy flux of gravitational radiation involving an ECO is different from the case of black hole. This will induce a dephasing of GWs during inspiral stage and potentially can be detected. Naively, we can expect that a longer inspiral signal makes easier to distinguish the difference between a black hole and an ECO. The long inspiral asks a large mass ratio. For example, the extreme-mass-ratio binaries (EMRBs), composed by a small black hole (only a few solar masses), neutron star (NS) or even primordial black hole (PBH) and a much massive  black hole (MBH, from a few tens to more than 100 solar masses ), will be ideal sources for the constraint of ECOs.  

We will see that for the EMRB with mass ratio 60:1, the influence of the quantum surface of ECO on the energy flux will induce detectable depahsing on the insprial signal. LVK has found several events with obvious asymmetric mass-ratio. GW190814 has two components with 23 and 2.6 solar masses; GW200210\_092254 is composed by 24 and 2.8 solar mass objects. Especially, GW191219\_163120 has the largest mass-ratio until now, 31 : 1.2. In the coming O4 run of LIGO-Virgo-KAGRA, it is possible to detect more large mass-ratio bianries, i.e., EMRBs.

This paper is organized as follows. In the next section, EMRB and detectability are discussed. In Section 3, we introduce our numerical method for calculating the waveforms. The energy flux and the result will be described in the Section 4. Finally, we will make a conclusion.

\section{EMRBs} 

\begin{figure}
\centering
\includegraphics[height=2.0in]{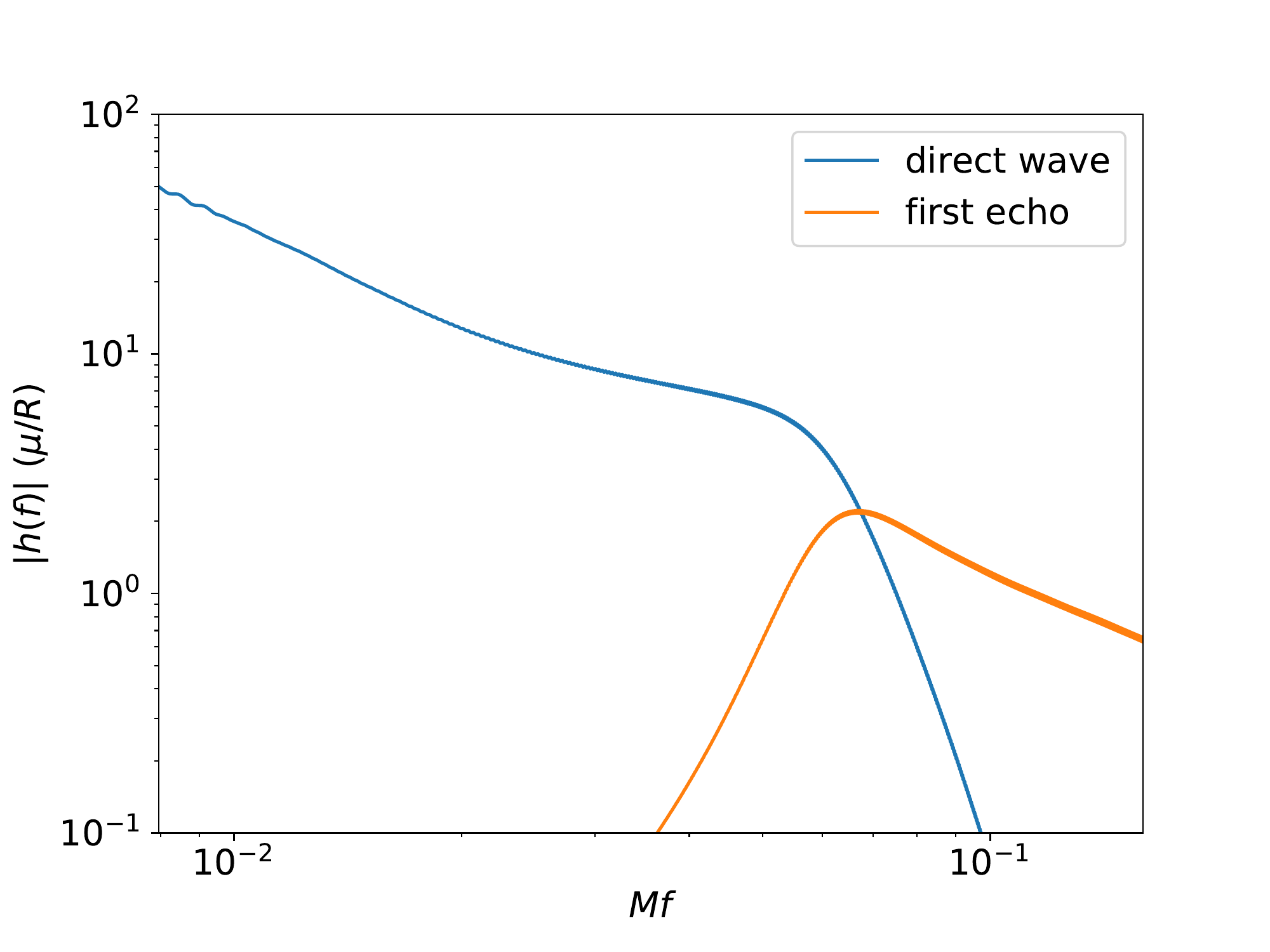}
\caption{frequency domain signals of inspiral waveform (blue solid) and echo (orange solid).}  \label{gwecho}
\end{figure}

\begin{figure}
\begin{center}
\includegraphics[height=2.0in]{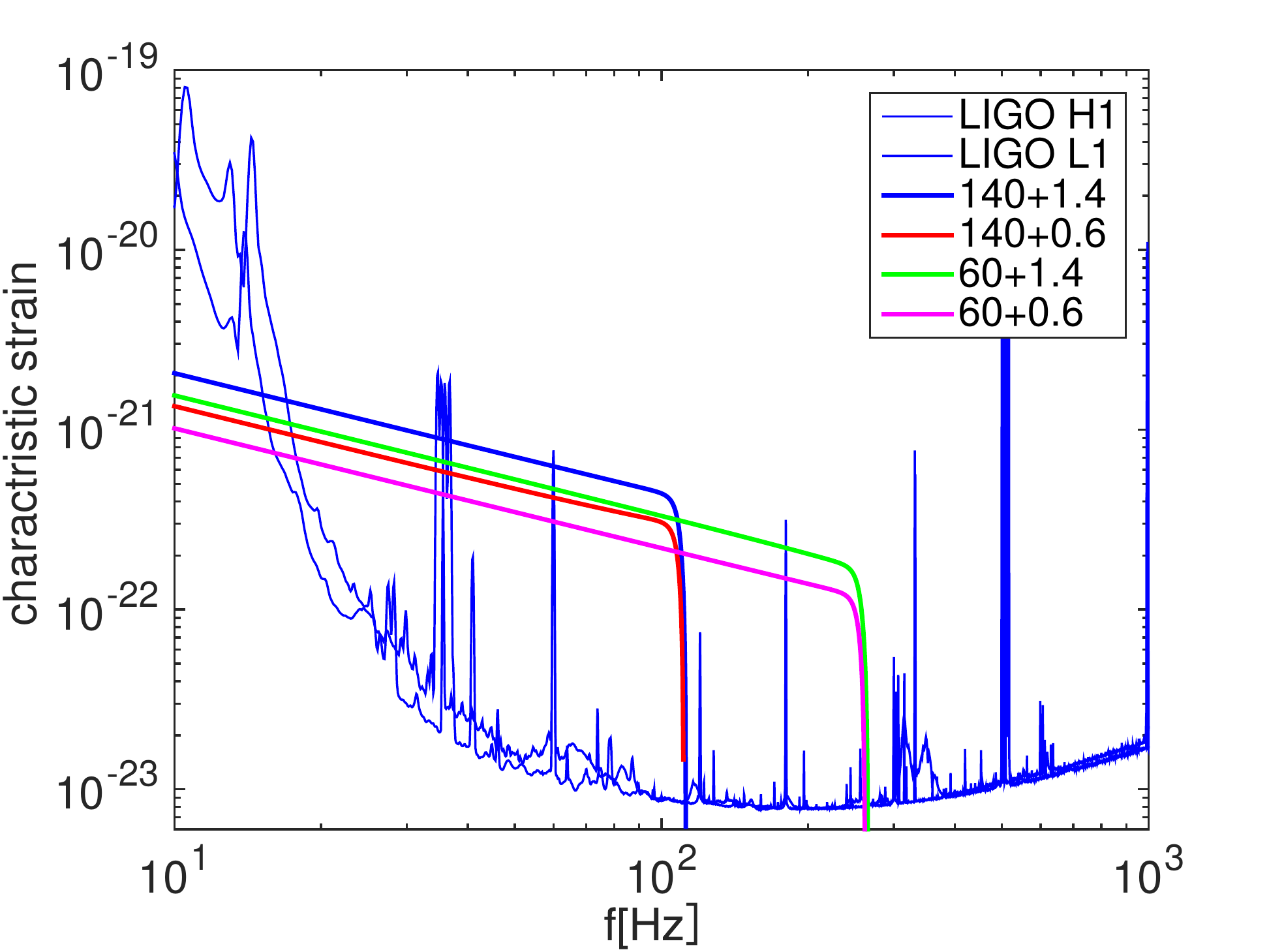}
\caption{Characteristic strains of four EMRBs composed by massive black holes (140 or 60 solar masses) and neutron stars (1.4 $M_\odot$) or primordial black holes (0.6 $M_\odot$), the sources are located at 100 Mpc from the Earth.}  \label{imris}
\end{center}
\end{figure}

Up to now, LIGO-Virgo has detected the most massive black hole with  107 $M_\odot$ before merger, and the final mass of the remnant black hole at 174 $M_\odot$ in GW170426\_190642 event \citep{gwcatalog}. The lightest compact object (neutron star) is 1.2 $M_\odot$ in GW191219\_163210. It is worth to expect
LIGO-Virgo-KAGRA will detect extreme-mass-ratio binaries composed by a very heavy black hole and a light compact object with mass ratio typically larger than 20 in the future. This kind of sources has enough signal to noise ratio (SNR) for detections (Fig. \ref{imris}), even the second small objects are neutron stars (NSs) with 1.4 $M_\odot$ or primordial black holes (PBHs) with only 0.6  solar mass. In this work, we use 4 EMRBs: NS (1.4 solar mass) + 140 solar mass MBH/ECO (Sys 1), PBH (0.6 solar mass) + 140 solar mass MBH/ECO (Sys 2),  neutron star + 60 solar mass MBH/ECO (Sys 3), and PBH + 60 solar mass MBH/ECO (Sys 4) to demonstrate the potential capability of LIGO-Virgo constraint on the ECOs. For calculating the gravitational waves from the small compact objects inspiraling into the massive BHs or ECOs, especially the effects due to hard surface of ECOs, we employ the Teukolsky equation \citep{teukolsky73} to fulfill this task by introducing ``reflectivity" of the ECO surface.  Distinguishing with the previous works of searching the direct echoes (i.e., the orange line in Fig. \ref{gwecho} )after ringdown signals, we focus on the influence of the surface on the inspiraling part before plunge (i.e., the blue line in Fig. \ref{gwecho} ).   

\section{Waveforms and orbital evolution}

By using effective-one-body (EOB) dynamics and black hole perturbation theory, gravitational radiations from EMRBs with have been investigated for Schwarzschild in \citep{Bernuzzi10}  and Kerr MBHs \citep{han2011}. The later developed the so called ``ET" codes to use the Teukolsky-based fluxes source the EOB dynamics. After decomposing $\psi_4$ in a Fourier transformation $\psi_4=\rho^4\int^{+\infty}_{-\infty}{d\omega\sum_{lm}{R_{lm\omega}(r)_{~-2}S^{a\omega}_{lm}(\theta)e^{-i\omega t+im\phi}}}$ with $_{~-2}S^{a\omega}_{lm}$ the spin-weighted spheroidal harmonics, the radial master equation of the Teukolsky equation is
\begin{align}
\Delta^2\frac{d}{dr}\left(\frac{1}{\Delta}\frac{dR_{lm\omega}}{dr}\right)-V(r)R_{lm\omega}=-\mathcal{T}_{lm\omega}(r),
\label{radialmaster}
\end{align}
where $\mathcal{T}_{lm\omega}(r)$ is the source term, which is decided by the energy-momentum tensor of the perturber, and $V(r)$ is the potential. Due to the Teukolsky potential is long-ranged, making it hard to numerically extract certain parameters in homogeneous solution,  Sasaki and Nakamura transformed the radial equation so that the potential is short-ranged and the equation become numerical computable\citep{sasaki}, then the Sasaki-Nakamura equation for  function $X^{}_{lm\omega}$:
\begin{align}
\frac{d^2
	X_{lm\omega}}{dr*^2}-F(r)\frac{dX_{lm\omega}}{dr*}-U(r)X_{lm\omega}=0 \,,\label{s-keq}
\end{align}
where $r*$ is the tortoise coordinate. The functions $\alpha, \beta, \eta$ and the potentials $F(r), U(r)$ can be found in \citep{Hughes00}. 
Note that the expression for $\eta$ given in \citep{Hughes00} is an expansion over $1/r$, which fails to recover reasonable results when r goes to horizon (because $eta$ has to diverge as $\Delta^2$ when $r\rightarrow r_+$). Original expression for $eta$ is given in Eq. (2.8c) of Sasaki-Nakamura's original article \citep{sasaki}:
\begin{equation}
	\eta = \alpha (\alpha + \frac{d\beta/dr}{\Delta} ) -  \frac{\beta}{\Delta}(\frac{d\alpha}{dr} +\frac{\beta V(r)}{\Delta^2} )
\end{equation}

The Sasaki-Nakamura equation admits two homogeneous solution having the purely sinuous asymptotic behavior due to the short-rangeness of potential $U(r)$:
\begin{equation}
\begin{aligned}
\label{eq_XH}
X^{H}_{lm\omega}&=A^{hole}_{lm\omega}e^{-ipr*},\quad r\rightarrow r_+,\\
X^{H}_{lm\omega}&=A^{\rm{out}}_{lm\omega}e^{i\omega
	r*}+A^{\rm{in}}_{lm\omega}e^{-i\omega r*},\quad r\rightarrow
\infty;
\end{aligned}
\end{equation}
and
\begin{equation}
\begin{aligned}
\label{eq_Xinf}
X^{\infty}_{lm\omega}&=C^{\rm{out}}_{lm\omega}e^{ip
	r*}+C^{\rm{in}}_{lm\omega}e^{-ip r*},\quad r\rightarrow r_+,\\
X^{\infty}_{lm\omega}&=C^{\infty}_{lm\omega}e^{i\omega r*},\quad r\rightarrow
\infty,
\end{aligned}
\end{equation}
Now, an ECO has a reflecting boundary at $r_0$ where is very near the position of horizon $r_+$, i.e.,  $r_0 = r_+ + \epsilon$, and $\epsilon \ll 1$. The solution reflected by the boundary is the combination of $X^H $ and $X^\infty$
\begin{equation}
	X^{\rm ref}_{lm\omega} = \mathcal{K} X^\infty_{lm\omega} + X^H_{lm\omega}
\end{equation}
which satisfies the reflecting boundary condition at the surface $r_0 = r_+ + \epsilon$ with reflectivity $\tilde{\mathcal{R}}$
\begin{equation}
	X^{\rm ref} \propto e^{-ip(r*-r_0*) } +\tilde {\mathcal{R}} e^{ip(r*-r_0*)},\quad \quad r* \rightarrow r_0* \,. \label{xsolution}
\end{equation}

Taking the asymptotic behavior of $X^{H,\infty}_{lm\omega}$ in Eq. (\ref{eq_XH}, \ref{eq_Xinf}) into the above boundary condition, one can find $\mathcal{K}$:
{
\begin{equation}
\label{K}
	\mathcal{K} = \frac{\tilde{\mathcal{R}}  e^{-2ipr_0*} \mathcal{T}_{BH} }{1- \tilde{\mathcal{R}} e^{-2ipr_0*} \mathcal{R}_{BH}} \quad {\rm{ with }}  \quad \mathcal{T}_{BH}\equiv\frac{C^{\infty}}{C^{out}}, \, \mathcal{R}_{BH} \equiv \frac{C^{in}}{C^{out}} \,.
\end{equation}
}

Finally, the GW waveform from a small object inspiraling into an ECO is
\begin{equation}
\begin{aligned}
&h^{\rm ECO}_+(R,\Theta,\Phi,t)-ih^{\rm ECO}_\times(R,\Theta,\Phi,t)=\\&\frac{2}{R}\sum_{lm}\frac{1}{\omega^2}(Z^H_{lm\omega}+
\mathcal{K}\frac{D^{\infty}_{lm\omega}}{B^{\rm hole}_{lm\omega}}Z^{\infty}_{lm\omega} )
\,_{-2}S^{a\omega}_{lm}(\Theta)e^{i[m\phi-\omega_m(t-r*)]}.
\label{waveform}
\end{aligned}
\end{equation}
The energy fluxe of gravitational radiation is decomposed by two parts $\dot{E} = \dot{E}^\infty+\dot{E}^{\rm H}$ (here the dot means $d/dt$), i.e., the flux goes to the infinity and down to the horizon. Where, the infinity flux is written as
\begin{equation}
	\dot{E}^\infty =\sum_{l,m}  \frac{|Z^H_{lm\omega} + \mathcal{K}\frac{D^\infty_{lm\omega}}{B^{\rm hole}_{lm\omega}} Z^\infty_{lm\omega}|^2}{4\pi \omega^2} \,, \label{E8dot}
\end{equation}
and the horizon one is
\begin{equation}
\dot{E}^{\rm H} =\sum_{l,m} \alpha_{lm\omega} \frac{|{\mathcal{T}} Z^\infty(1+\mathcal{K}\frac{D^{\rm in}}{B^{\rm hole}})|^2}{4\pi \omega^2} \,. \label{EHdot}
\end{equation}
The transmissivity ${\mathcal{T}}$ has a relation with reflection factor: ${\mathcal{T}}^2 + {\mathcal{R}}^2 = 1$.  The relation between asymptotic amplitudes of Sasaki-Nakamura functions and Teukolsky functions \citep{sasakireview}:
\begin{equation}
\label{trans_AB}
\begin{aligned}
	B^{\rm in}_{lm\omega} = -\frac{1}{4\omega^2} A^{\rm in}_{lm\omega}, \quad
	B^{\rm hole}_{lm\omega} = -\frac{1}{d_{lm\omega}} A^{\rm hole}_{lm\omega}\\
\end{aligned}
\end{equation}
\begin{equation}
\begin{aligned}
D^{\rm in}_{lm\omega} =& -\frac{1}{d_{lm\omega}} C^{\rm in}_{lm\omega},\quad D^{\infty}_{lm\omega} =& -\frac{4\omega^2}{c_0} C^{\infty}_{lm\omega}\\
\end{aligned}\label{SN_D}
\end{equation}
The coefficients $c_0$, $d_{lm\omega}$ and $\alpha_{lm\omega}$ can be found in \citep{Hughes00}. Naturally, if we set $R = 0$, all the results go back the the familiar equations of black holes.

\section{Energy fluxes} 

Firstly, due to the existence of hard surface, the total energy fluxes will be slightly different from the case of black hole. How to calculate the energy fluxes is expressed in Eqs. (\ref{E8dot},\ref{EHdot}). If we choose $\mathcal{K} = 0$, i.e., $\tilde{\mathcal{R}} = 0$, then the energy fluxes will go back the case of black hole. The difference of total energy fluxes between ECO and BH cases depends on the orbital radius $r$ or equivalent the orbtial frequency $\omega_{\rm orb}$, reflectivity factor $\tilde{\mathcal{R}} = 0$, the location of the hard surface $r_0$ and also the spin of central body $a$. It is easy to find from Eqs. (\ref{xsolution},\ref{K}) that the relation of $\dot{E}$ with $r_0^*$ is just a sine or cosine function because of the $r_0^*$ is only in the exponential index.

\begin{figure}
\centering
\includegraphics[width=0.49\linewidth,scale=1]{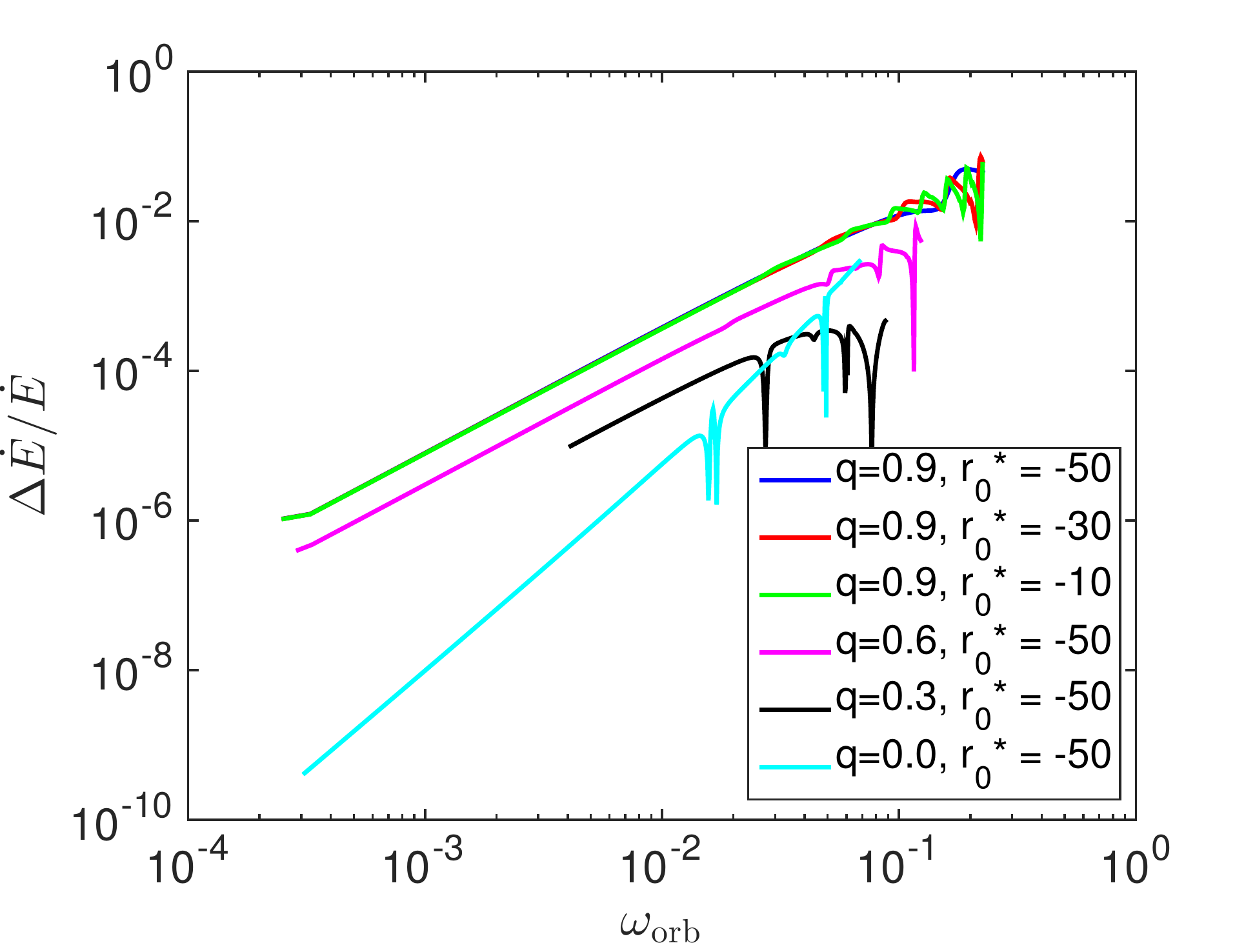}
\includegraphics[width=0.49\linewidth,scale=1]{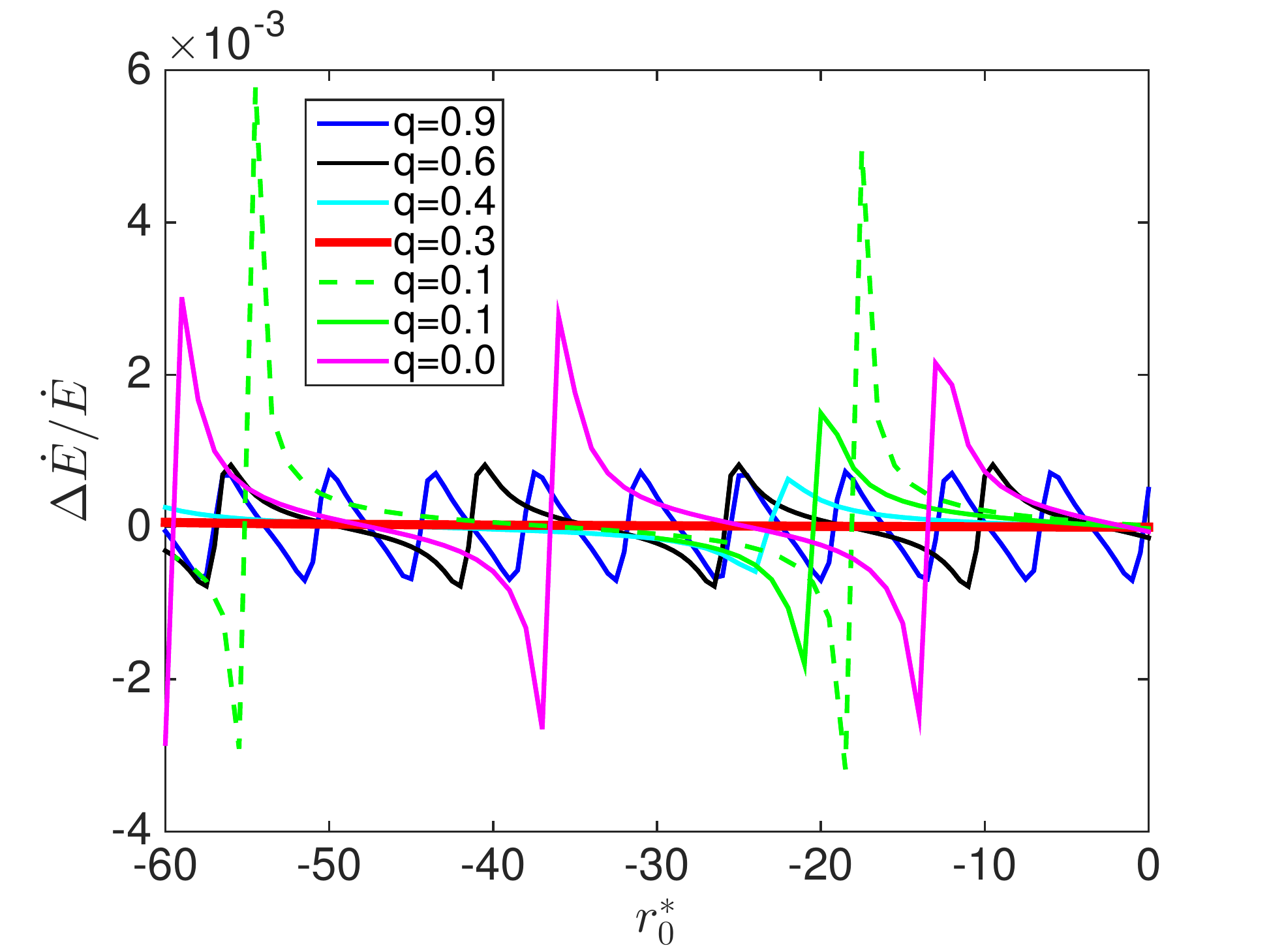}
\caption{Relative differences of energy fluxes $\Delta \dot{E}/\dot{E}$ between ECO and BH. Left panel: $\Delta \dot{E}/\dot{E}$ as a function with varied orbital frequency ; right panel: $\Delta \dot{E}/\dot{E}$ as a function with varied suface location for different spins.}  \label{dEdtvswr0}
\end{figure}

Fig. \ref{dEdtvswr0} shows the relative differences of energy fluxes $\Delta \dot{E}/\dot{E}$ between ECO and BH when one changes the orbital frequency and the location of ECO surface. We can clearly see that $\Delta \dot{E}/\dot{E}$ have a clear power-law with the orbital frequency when the frequency is not large. This phenomenon has been revealed by the post-Newtonian approximation \citep{Li2008}.  In the right panel of this figure, we can find that when $a = 0.3$, the difference of energy fluxes is almost zero, this is due to the orbital frequency is almost equal to the horizon frequency of the central body. This will be discussed later. 

\begin{figure}
\centering
\includegraphics[width=0.49\linewidth,scale=1]{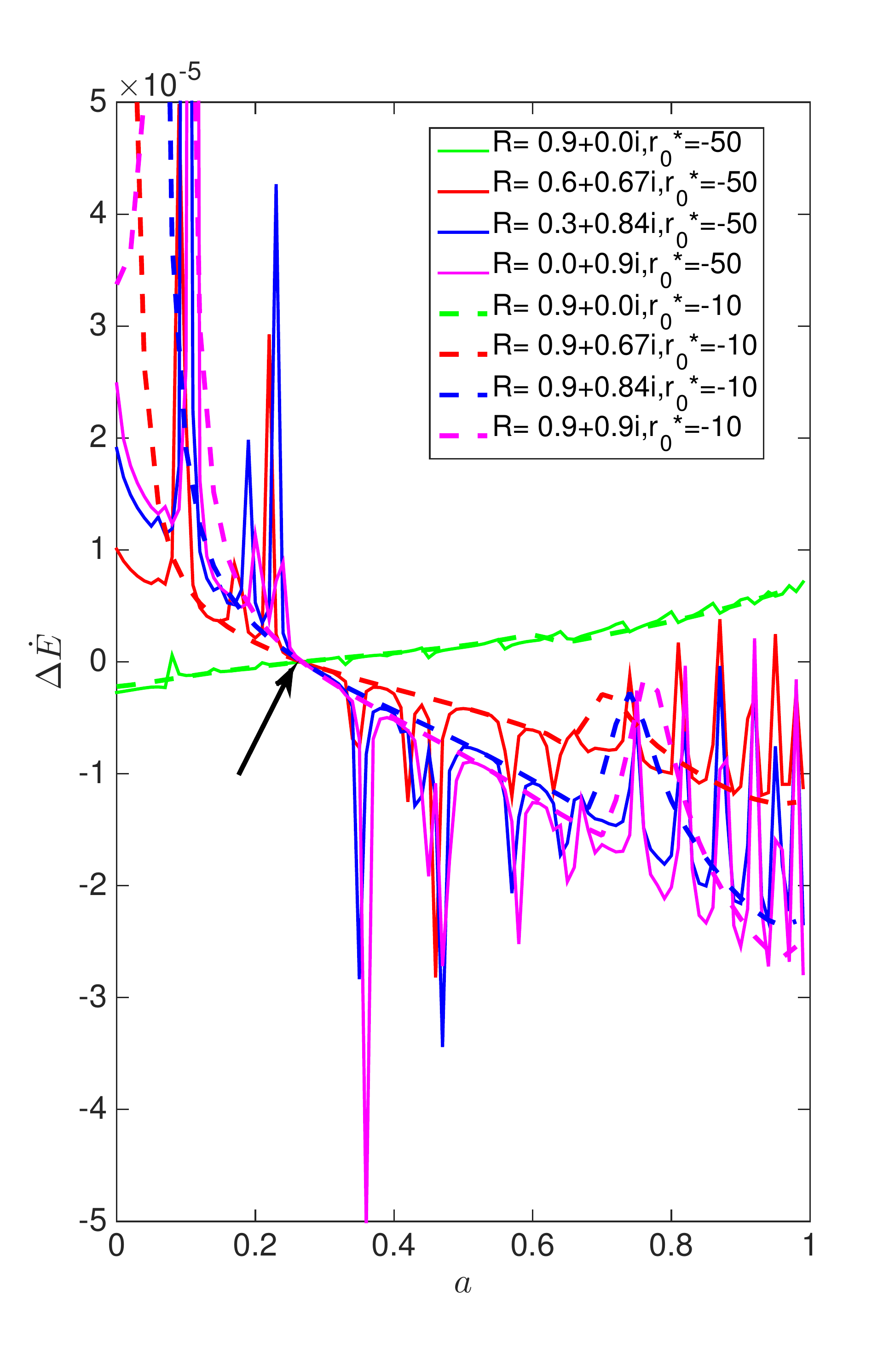}
\includegraphics[width=0.49\linewidth,scale=1]{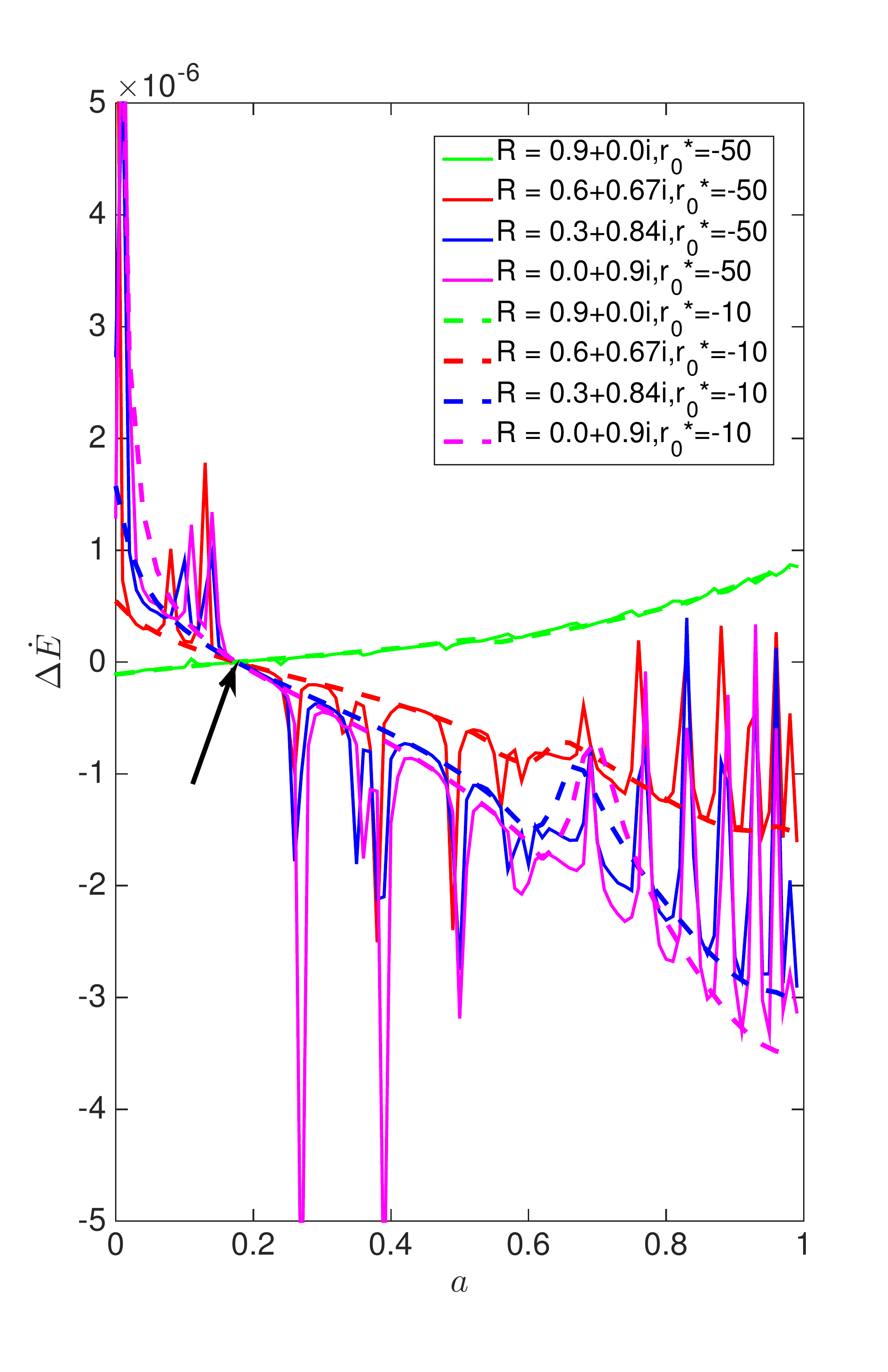}
\caption{The differences of energy fluxes $\Delta \dot{E}$ between ECO and BH. Left panel: $\Delta \dot{E}$ as a function versus spin with varied phase of reflectivity factors, and the orbital frequency is fixed to the ISCO one for nonspinning BH or ECO; right panel: The same with the left panel but the orbital frequency is fixed to the one with orbital radii $r = 8 M$ for non-spinning BH or ECO.}  \label{dEdtvsa_phase}
\end{figure}

Now, let's see how the differences of energy fluxes varying with different spin.  In Fig. \ref{dEdtvsa_phase}, for different phases of reflectivity factors, the magnitude and oscillation are different. Very interestingly, all the lines are gathered at zero for a special spin value. We find that at this spin value, the frequency of BH horizon or the quantum surface of ECO equals to the orbital frequency.

We can look the details how the energy fluxes varies with the phase of reflectivity factor, see Fig. ref{dEdtvsphase}.
\begin{figure}
\centering
\includegraphics[width=0.49\linewidth,scale=1]{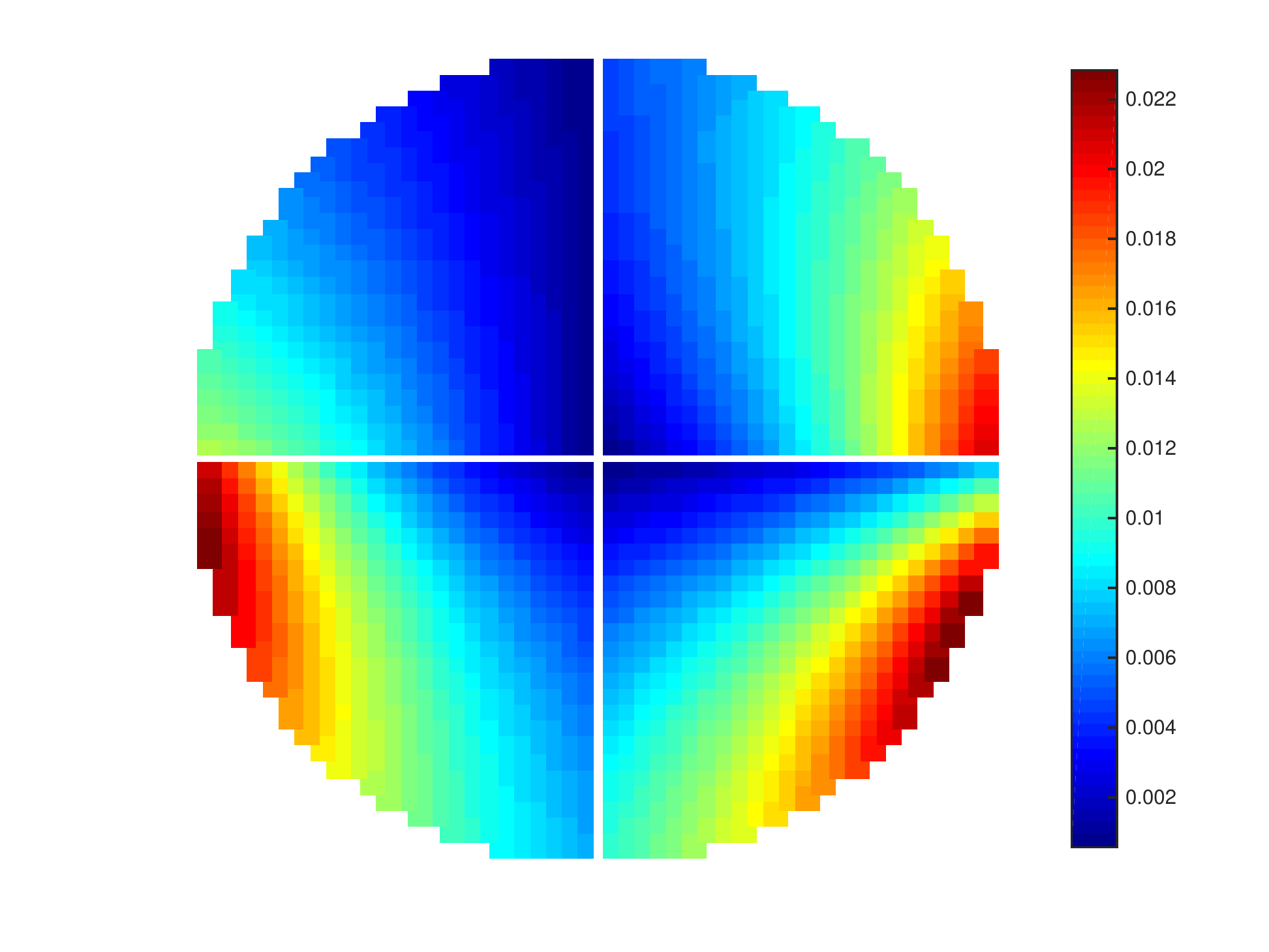}
\includegraphics[width=0.49\linewidth,scale=1]{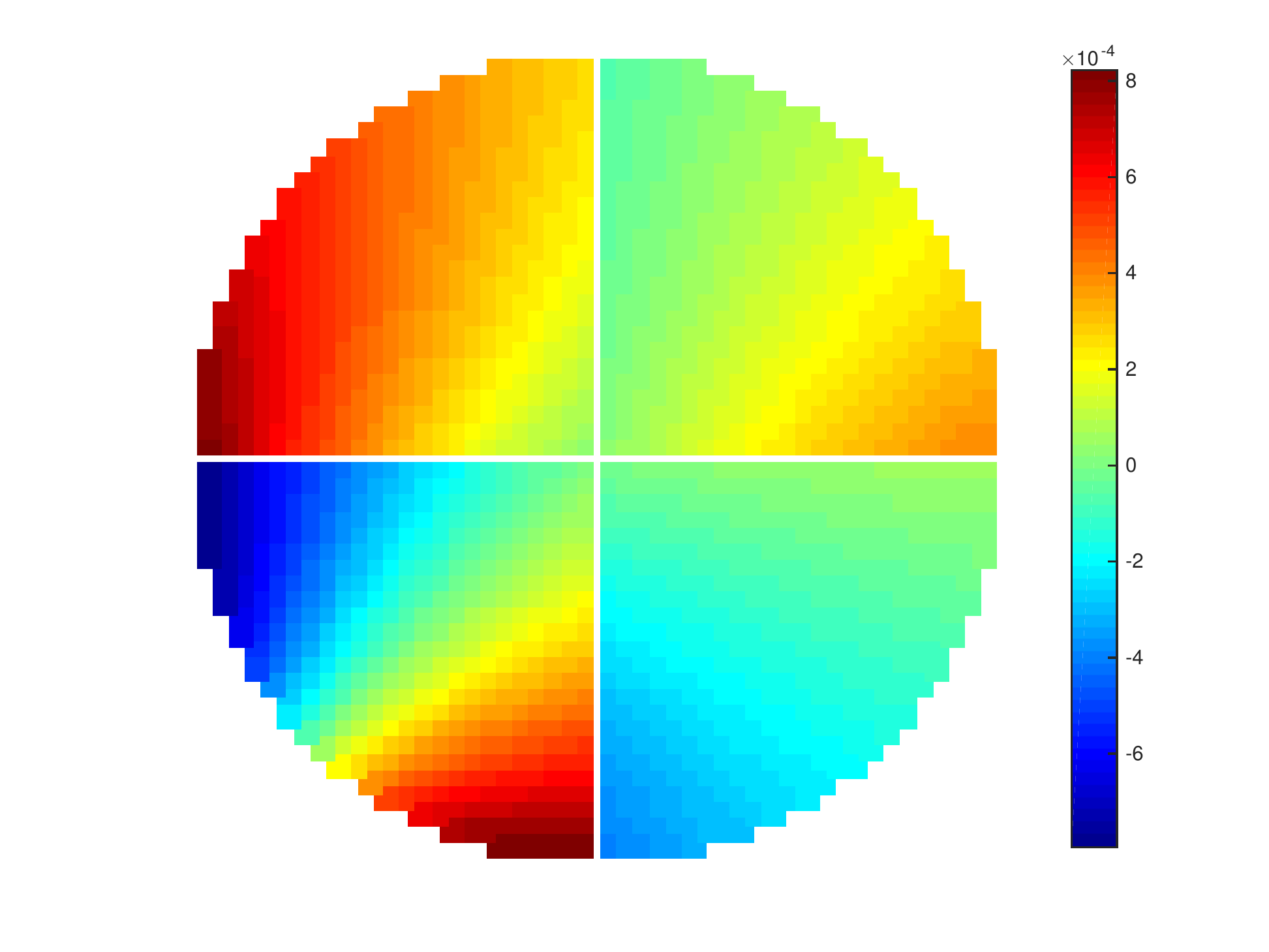}
\caption{Relative differences of energy fluxes $\Delta \dot{E}/\dot{E}$ between ECO and BH. Left panel: $\Delta \dot{E}$ of a small object orbiting at radii 3 $M$ around a BH or ECO with spin 0.9 as a function versus varied phase of reflectivity factors, the location of ECO surface are -50, -30, -20 and -10 from top-right panel to bottom-right panel counterclockwisely ; right panel: The same with the left panel but the orbital radii $r = 6 M$.}  \label{dEdtvsphase}
\end{figure}

\begin{figure}
\centering
\includegraphics[width=0.49\linewidth,scale=1]{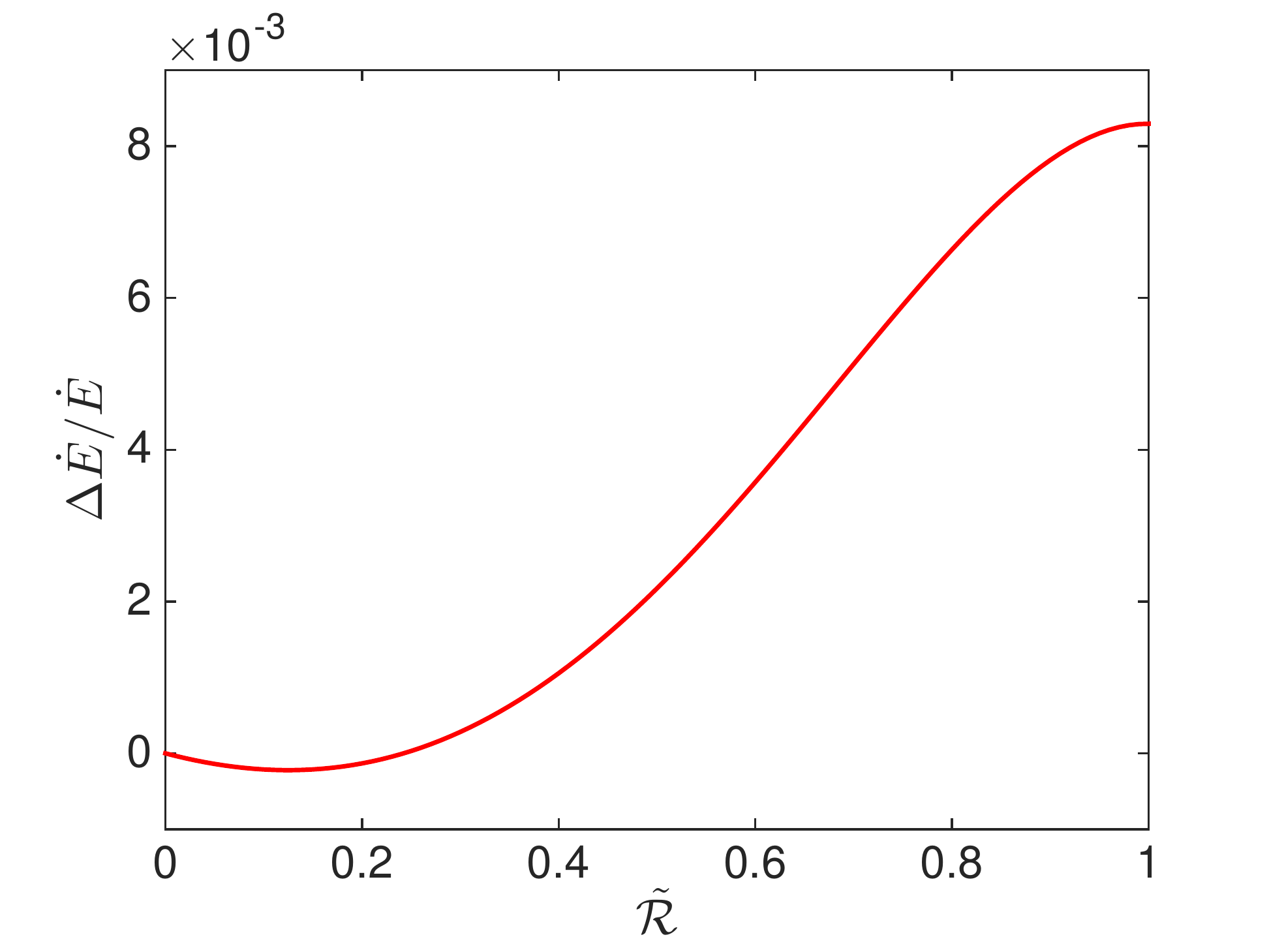}
\includegraphics[width=0.49\linewidth,scale=1]{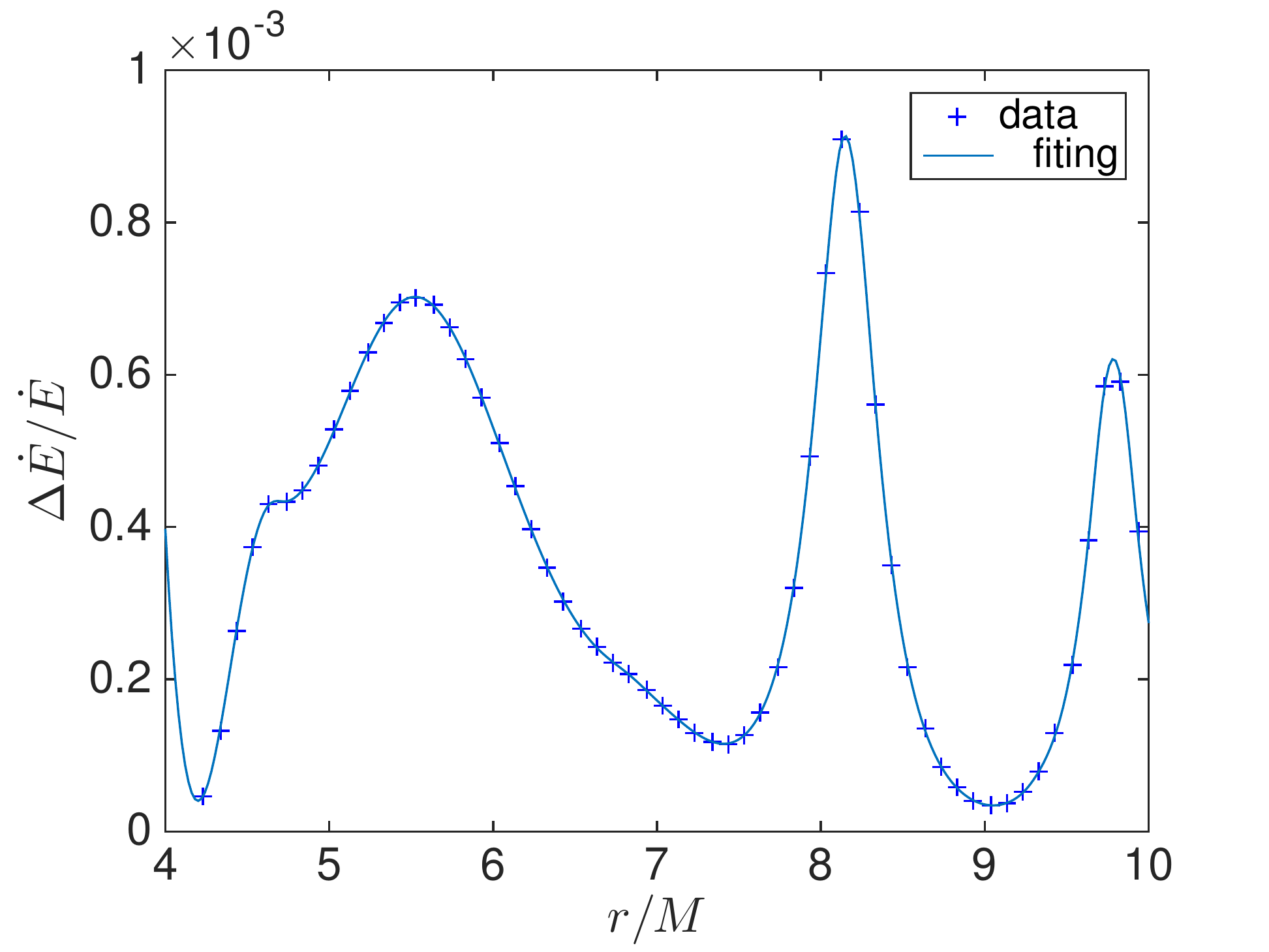}
\includegraphics[width=0.49\linewidth,scale=1]{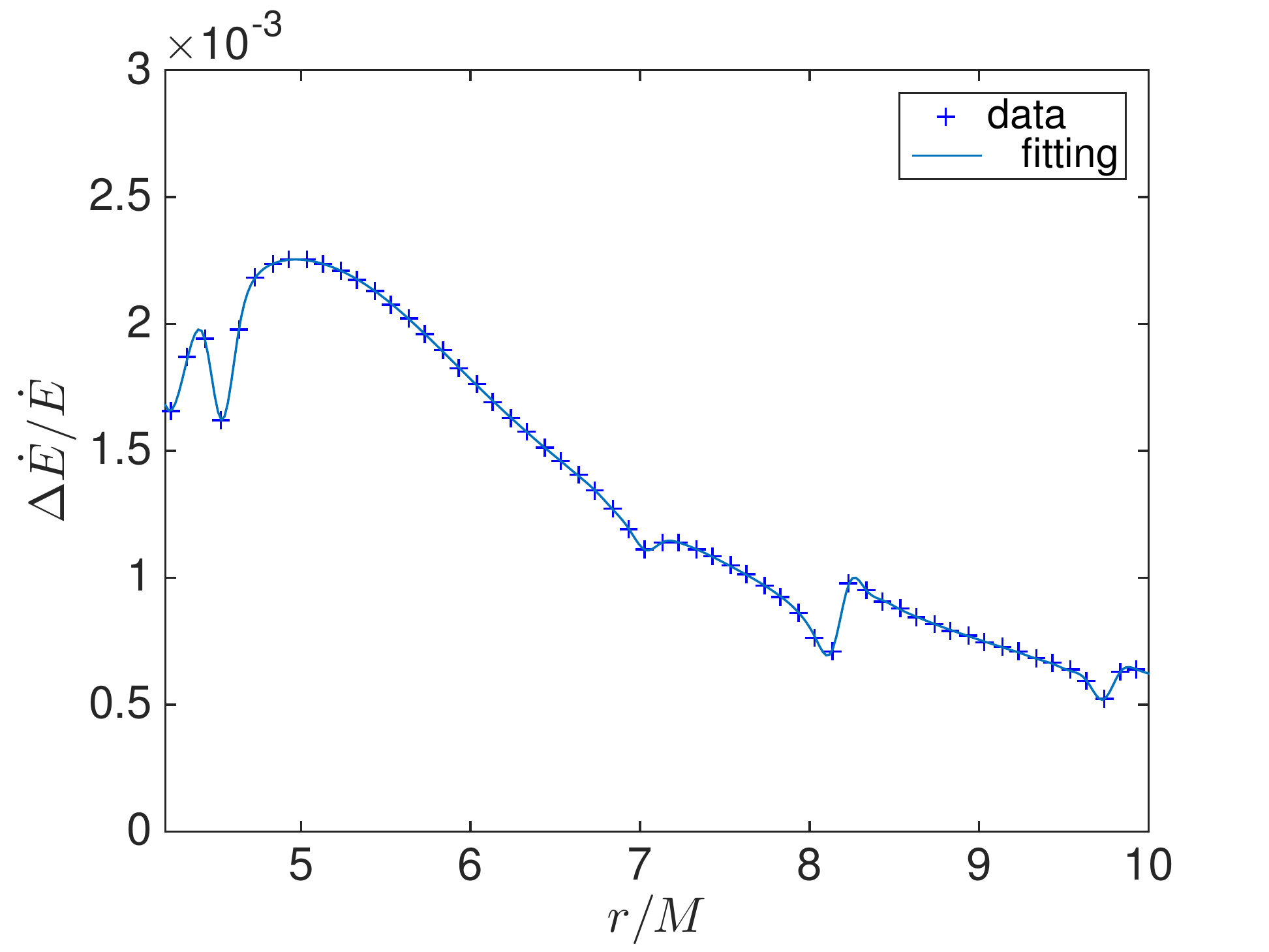}
\includegraphics[width=0.49\linewidth,scale=1]{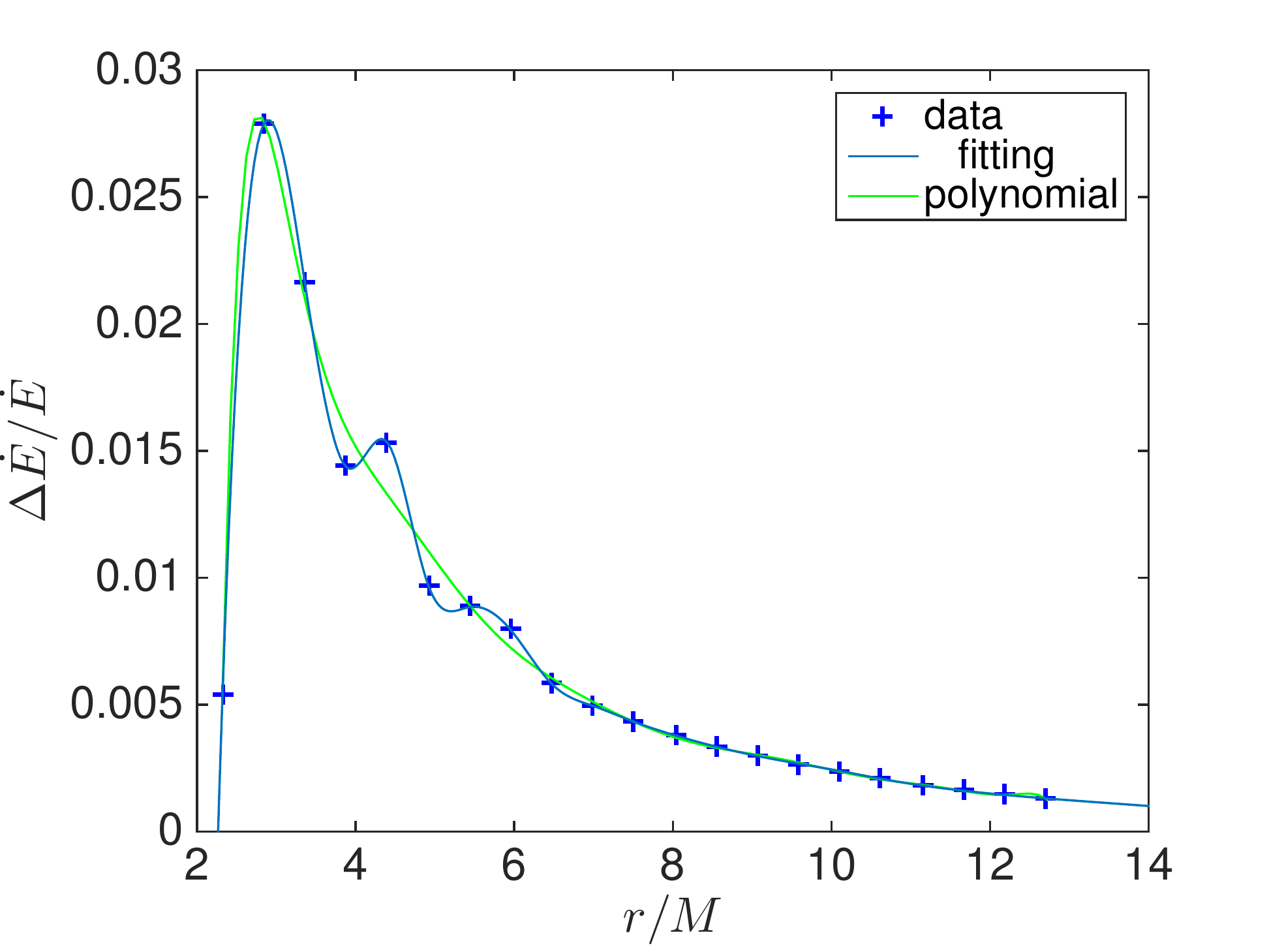}
\caption{Relative differences of energy fluxes $\Delta \dot{E}/\dot{E}$ between ECO and BH. Top-left panel: $\Delta \dot{E}/\dot{E}$ as a function with varied reflectivity $\tilde{\mathcal{R}}$ when $a = 0.9, r_0^* = -50$ and $r = 6 M$; top-right panel: $\Delta \dot{E}/\dot{E}$ as a function with varied orbital radii when $a = 0.5, r_0^* = -50$ and $\tilde{\mathcal{R}} = 0.5$; bottom panels: The same with the top-right one but $\tilde{\mathcal{R}} = 0.9$, $a = 0.5$ and 0.9 for left and right one respectively.}
\label{dEdt}
\end{figure}

Fig. \ref{dEdt} shows the relative differences of energy fluxes $\Delta \dot{E}/\dot{E}$ between ECO and BH when one changes the reflectivity factor and orbital radii. We can clearly see that the magnitude of  $\Delta \dot{E}/\dot{E}$ is at the order of $10^{-3} - 10^{-2}$. For comparable mass-ratio binaries, this error is too small to produce observable effect in LIGO band. However, if the mass-ratio becomes smaller, for example down to $10^{-2}$, this flux error should produce enough dephasing because of the much more cycles before merger and may induce a failure of detection. 

\begin{figure}
\centering
\includegraphics[width=0.59\linewidth,scale=1]{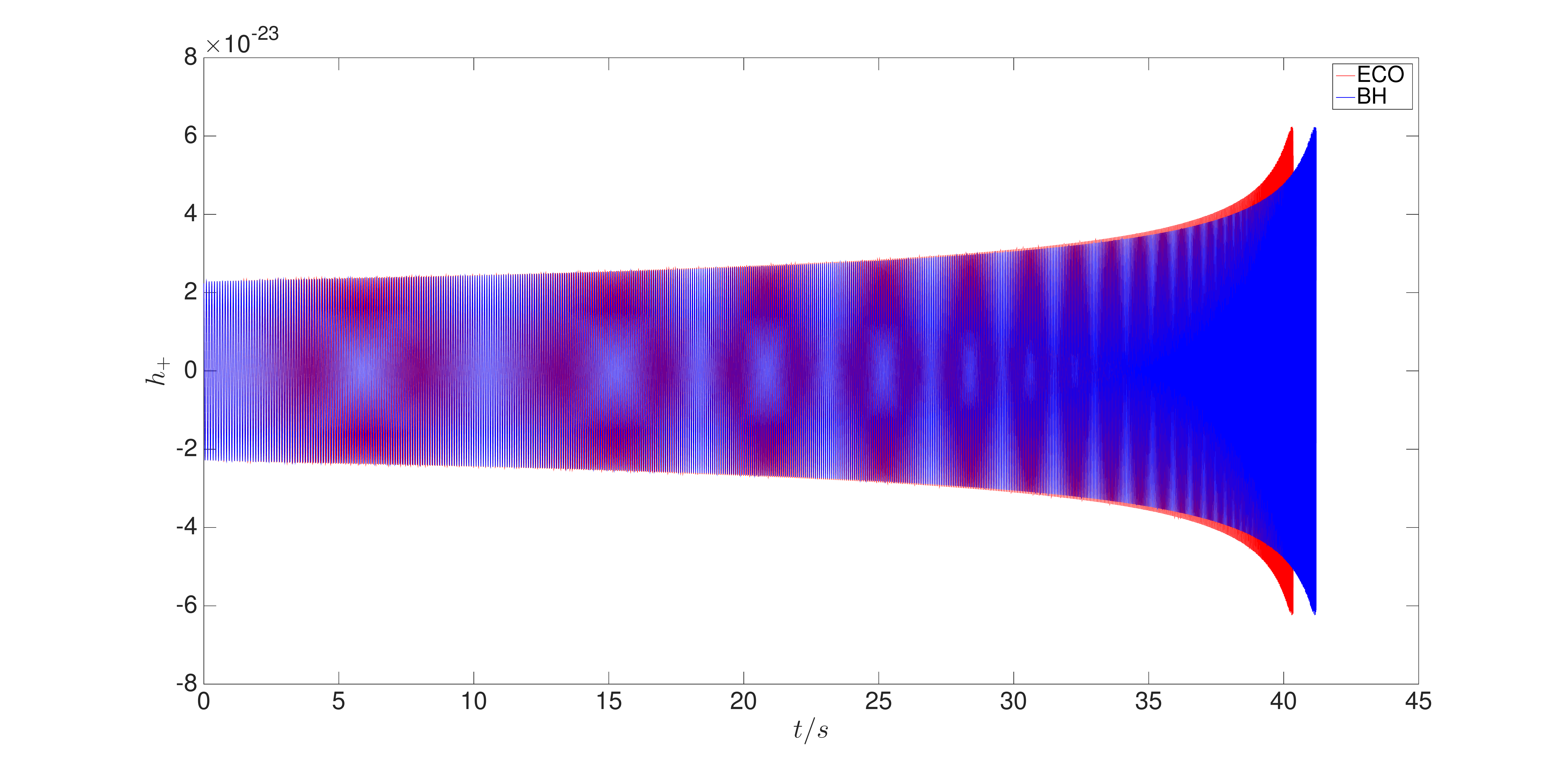}
\includegraphics[width=0.39\linewidth,scale=1]{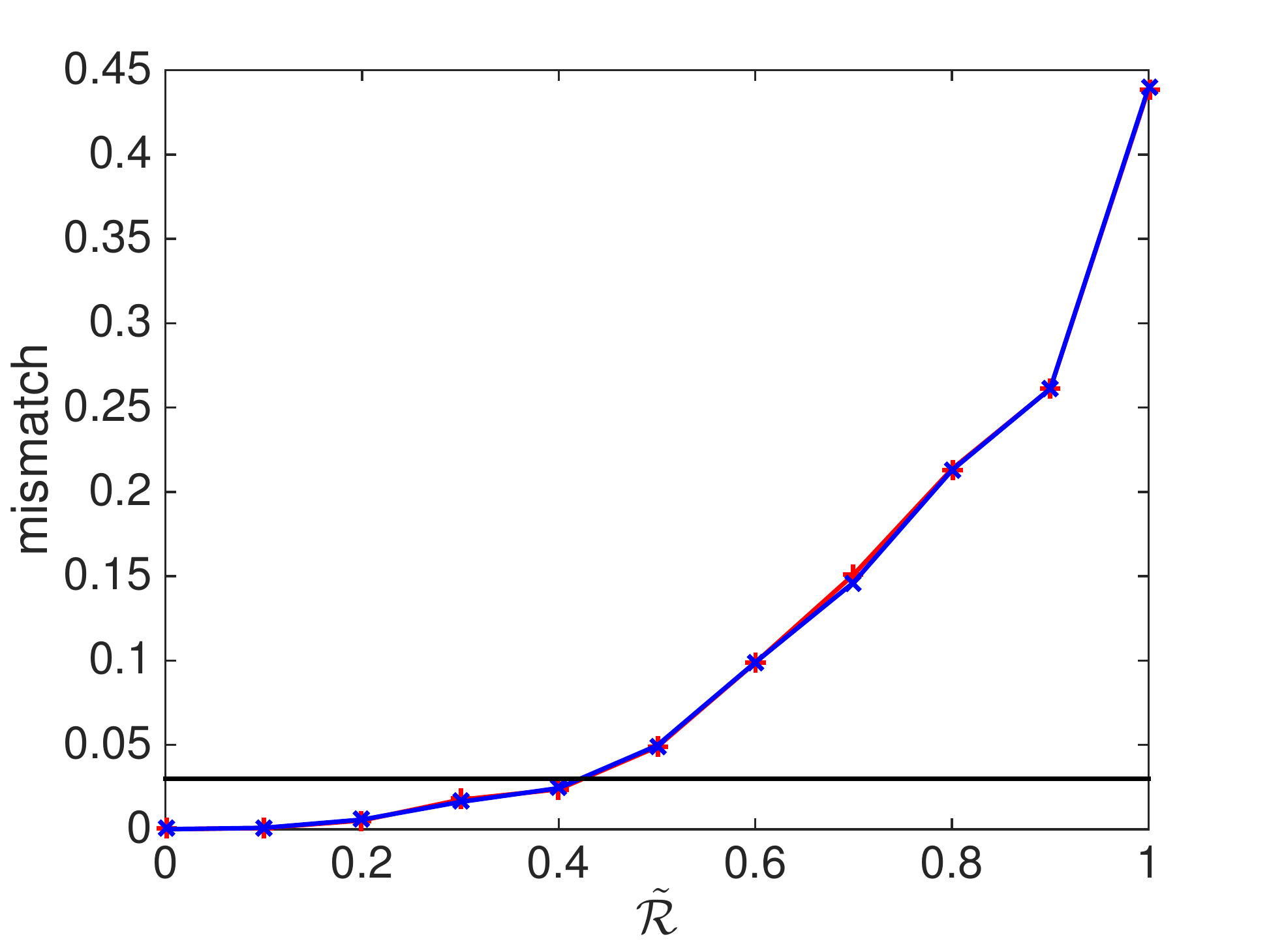}
\caption{Left panel: gravitational waves of MBH and ECO's EMRBs. The mass of massive compact body (ECO or BH) is 140 $M_\odot$ and the small object is 1.4 $M_\odot$, the spin of massive body is 0.9, reflectivity is 0.9 and surface location $r_0^* = -50$. Right panel: mismatch results of the two kind of waveforms while the reflectivity $\tilde{\mathcal{R}}$ of the hard surface of ECO from 0 to 1.} 
\label{waveforms}
\end{figure}

For investigating the effects on GWs from the hard surface of ECO, we use the  energy fluxes (\ref{E8dot},\ref{EHdot}) to evolve the orbits of EMRBs in Fig. \ref{imris} and generate the waveforms by Eq. (\ref{waveforms}). In the left panel of Fig. \ref{waveforms}, the waveform from a NS inspiraling into a MBH and the one from a NS inspiraling into an ECO with $\tilde{\mathcal{R}} = 0.9$ are shown. One can see an obvious difference between the two waveforms. By using matched filtering technology, we can quantitatively give out the difference of two waveforms ($h_a(t)$ and $h_b(t)$), i.e., mismatch. 
\begin{align}
    {\rm mismatch}  \equiv 1-\max \left(\frac{<h_a|h_b>}{\sqrt{<h_a|h_a><h_b|h_b>}}\right) \,,
\end{align}
where $<h_a|h_b>$ is the standard matched-filtering inner product between two waveforms,
\begin{align}
    <h_a|h_b> \equiv 2\int_0^\infty{df\frac{\tilde{h}_a^*(f)\tilde{h}_b(f)+\tilde{h}_a(f)\tilde{h}_b^*(f)}{S_n(f)}}\,,
\end{align}
where $\tilde{h}$ means the frequency-domain waveform and $*$ the complex conjugate. In the following analysis, the noise power spectral density$S_n(f)$ is taken from the LIGO noise. 
In the right panel of Fig. \ref{waveforms}, mismatch of waveforms from BH and ECO with varied $\tilde{\mathcal{R}}$ are shown. It demonstrates that once $\tilde{\mathcal{R}} > 0.5$, the hard surface of ECO will induce a considerable mismatch on the waveforms and in principle can be detected from the GW data.
% The \nocitep command causes all entries in a bibliography to be printed out
%\nocitep{*}

\begin{figure}
\begin{center}
\includegraphics[height=2.0in]{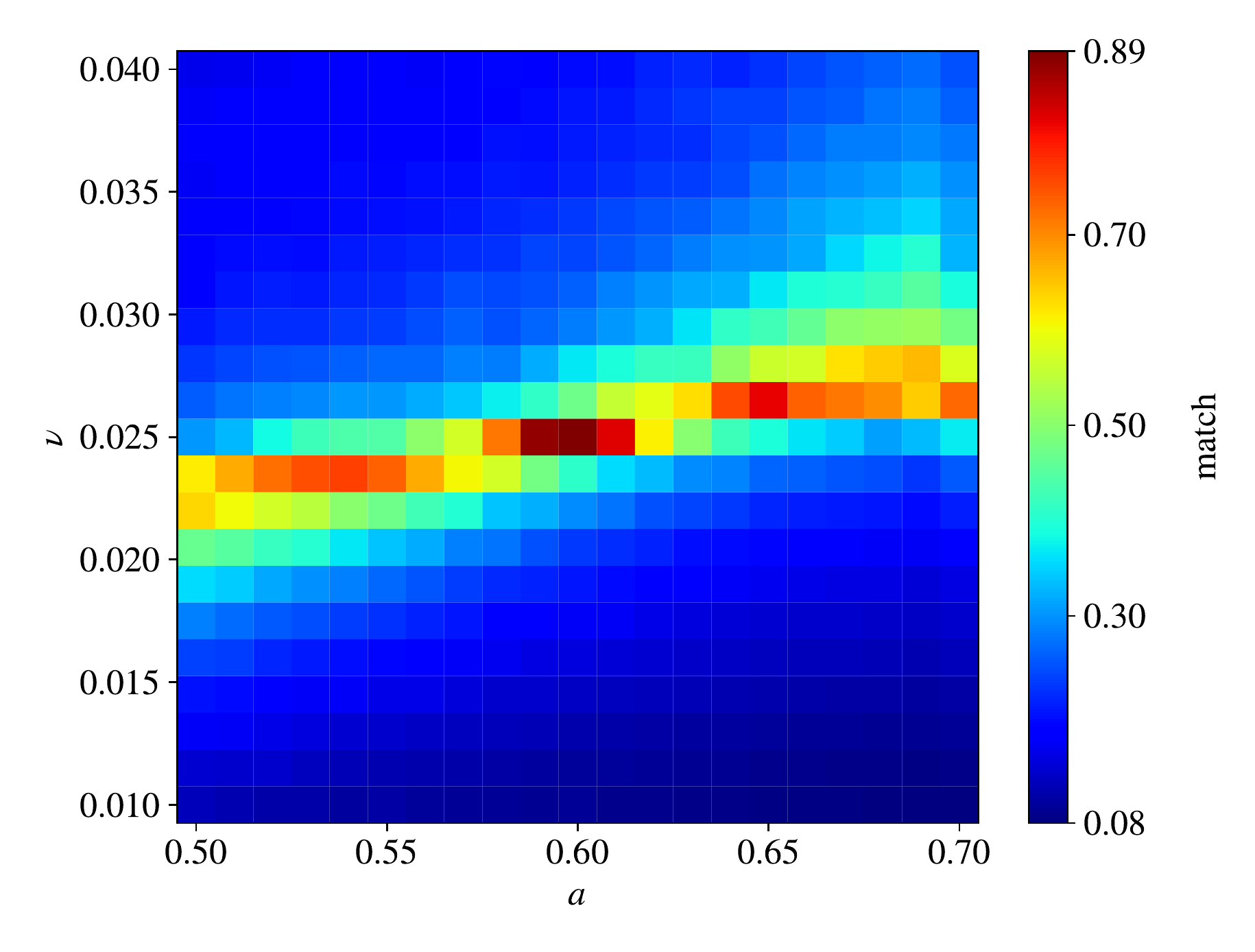}
\caption{match results between ECO waveforms and a group of BH waveforms with varied mass ratio and spin.}  \label{waveformmatch}
\end{center}
\end{figure}

However, one may suspect that the effect of hard surface on the energy fluxes can be mimicked by other physical parameters. The mass-ratio and spin $a$ can also change the energy fluxes and then the orbital evolution and waveforms. It means that the reflectivity is possibly degenerate with other parameters, and then can not be recognized from the GW data. For checking this point, we match the waveform of an EMRB 1.5 $M_\odot$ NS + 60 $M_\odot$ ECO  with $\tilde{\mathcal{R}} = 0.9$ and $a = 0.6$, and a group of EMRBs with mass-ratio from 0.01 to 0.04 and spin of BH from 0.5 to 0.7. In Fig. \ref{waveformmatch}, the match results show that the best match is the BH choose the same spin and mass-ratio of ECO. The best match is still less than 0.9, and for different spin and mass-ratio cases one can not find a waveform from BH can mimic the one from ECO. This may prove that the degeneracy of $\tilde{\mathcal{R}}$ does not exist. In other words, if the compact object is really an ECO, we can recognize it from BH and find the no-horizon extremely compact object. 

\begin{figure}
\begin{center}
\includegraphics[height=3.25in]{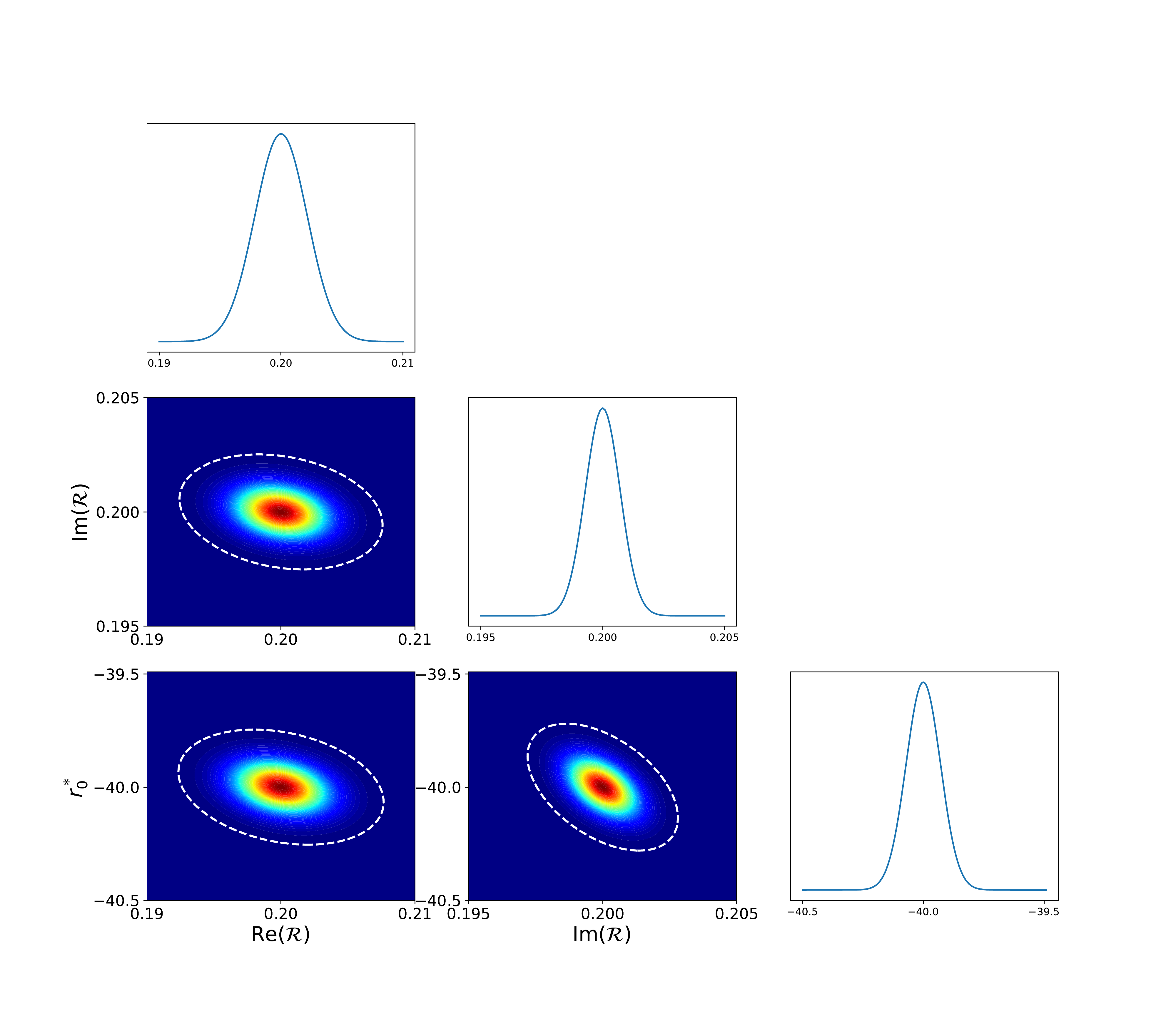}
\caption{Likelihood of $\mathcal{R}$ and $r^*_0$ which are derived from the Fisher matrix. The white dashed ellipse is shown at $3\sigma$.}  \label{likelihood}
\end{center}
\end{figure}

Furthermore, for estimating the measurement accuracy of the reflectivity factor and surface location by the
EMRB signals, we employ Fisher matrix to do the parameter estimation. We use
waveforms with the signal-to-noise ratio
(SNR) is about 130. The Fisher information matrix for a GW signal $h$
parameterized by $\boldsymbol{\lambda}$ is given by \citep{cutler94}

\begin{align}
\Gamma_{ij} = \left<\frac{\partial h}{\partial \lambda_i}|\frac{\partial h}{\partial \lambda_j}\right>\,,
\end{align}

\noindent
where $\boldsymbol{\lambda}$ is the waveform parameters including ${\mathcal R}$ and $r_0*$. In
the case of high SNR, the errors of parameters can be approximated as square
root of the diagonal elements of the inverse of $\Gamma_{ij}$, i.e., $\Delta \lambda_i \approx \sqrt{(\Gamma^{-1})_{ii}}$. We find that the
 reflectivity factor and surface location can be measured to about $\Delta \lambda_i/\lambda_i
\sim 1.3 \%$ and $1.5\%$ respectively. The corresponding likelihood is $\mathcal{L}({\boldsymbol\lambda}) \propto e^{-\frac{1}{2}\Gamma_{ij}\Delta\lambda_i\Delta\lambda_j}$ \citep{cutler94,babak17}, and is shown in Fig. \ref{likelihood}. These results demonstrate that once LIGO can catch EMRB events, without observing 
the echo signals, just from the inspiraling waveforms, one can constrain the properties of the surface of BHs in a positive accuracy.

\section{Conclusions} 

The nature of the extremely compact object is quite important in the theoretical physics. Though black holes are very popular, the other candidates like as firewall model, Boson star are still living \citep{cardoso19}. Due to the hard surface at the Plank scale near the corresponding position of BH horizon, there are a lot of prediction of GW echoes after ringdown in literature.  In 2017, Abedi et. al. claimed that they had found tentative evidence of echoes at a combined 2.9$\sigma$ significance level \citep{Abedi17a}. However, their result  was questioned \citep{Ashton17,Abedi17b}. Various teams have also proposed methods to estimate the parameters of the gravitational-wave echoes and to search for echoes \citep{echoSearch1,echoSearch2,echoSearch3,echoSearch4,echoSearch5,echoSearch6,echoSearch7}. Until now, there are no more reports about finding echo signals.

In the present paper, we do not consider the direct echo signals, alternatively, we try to reveal the effect of the quantum surface at Planck scale outside of horizon in the inspiraling stage of binary merger. Due to the exist of   
quantum surface, the energy fluxes and gravitational waveform are both modified. However, this modification is usually too small to be recognized in the GW data from comparable mass-ratio binaries. Fortunately, if the mass-ratio smaller than 1:10, thanks to the more long inspiral, this modification due to the possible hard surface will be extracted from the GW data.

By modifying the boundary condition of the Teukolsky equation, we numerically calculate the orbital evolution and waveforms from a compact object (NS, BH, PBH) inspiraling a massive ECO. Our calculation shows that if the hard surface really exists and its reflectivity factor is large enough ($\gtrsim 0.5$), by matched-filtering technology, we may have a chance to study the nature of extremely compact objects, and answer if they are black holes or ECOs.

\section*{Acknowledgements}
This work is supported by The National Key R\&D Program
of China (No. 2021YFC2203002), NSFC No. 11773059 and No. 12173071,  and the Strategic Priority Research Program of the CAS under Grants No. XDA15021102. W. H. is supported by CAS Project for Young Scientists in Basic Research YSBR-006. We thank Yanbei Chen, Shuo Xin, Ling Sun for very useful discussions. 

\section*{DATA AVAILABILITY}
The data underlying this article will be shared on reasonable request to the corresponding author Wen-Biao Han.


\end{document}